\documentclass[twocolumn]{emulateapj}
\journalinfo{{\sc Accepted by The Astrophysical Journal}}
\usepackage{natbib}
\usepackage{amssymb}
\usepackage{graphicx}
\newcommand{\igr}{IGR J16393-4643}

\begin{document}
\title{Orbital Parameters for the X-ray Pulsar IGR J16393-4643}
\author{Thomas W. J. Thompson\altaffilmark{1},
John A. Tomsick\altaffilmark{1},
Richard E. Rothschild\altaffilmark{1},
J. J. M. in 't Zand\altaffilmark{2}, 
Roland Walter\altaffilmark{3}}

\altaffiltext{1}{Center for Astrophysics and Space Sciences, University of California, San Diego, La Jolla, CA 92093; email: tthompson@physics.ucsd.edu}
\altaffiltext{2}{Space Research Organization of the Netherlands, Sorbonnelaan 2, 3584 CA Utrecht, Netherlands}
\altaffiltext{3}{{\em INTEGRAL} Science Data Centre, Chemin d'Ecogia 16, 1290 Versoix, Switzerland}

\begin{abstract}
With recent and archival {\em Rossi X-Ray Timing Explorer} ({\em RXTE}) X-ray measurements of the heavily obscured X-ray pulsar IGR J16393-4643, we carried out a pulse timing analysis to determine the orbital parameters. Assuming a circular orbit, we phase-connected data spanning over 1.5 years. The most likely orbital solution has a projected semi-major axis of 43 $\pm$ 2 lt-s and an orbital period of 3.6875 $\pm$ 0.0006 days. This implies a mass function of 6.5 $\pm$ 1.1 M$_{\sun}$ and confirms that this {\em INTEGRAL} source is a High Mass X-ray Binary (HMXB) system. By including eccentricity in the orbital model, we find $e < 0.25$ at the 2$\sigma$ level. The 3.7 day orbital period and the previously known $\sim$910 s pulse period place the system in the region of the Corbet diagram populated by supergiant wind accretors, and the low eccentricity is also consistent with this type of system. Finally, it should be noted that although the 3.7 day solution is the most likely one, we cannot completely rule out two other solutions with orbital periods of 50.2 and 8.1 days. 

\end{abstract}
\section{Introduction}
\igr~was initially discovered by Sugizaki et al. (2001) and is listed as AX J16390.4-4642 in the {\em ASCA} Faint Source Catalog. It was later detected by {\em INTEGRAL} during the first Galactic Plane Scan (GPS) of the Norma spiral arm (Bodaghee et al. 2006). The {\em INTEGRAL} GPSs have revealed many highly-absorbed sources that were not easily detectable with the soft bandpasses ($\la 10$ keV) of most previous observatories. The brightest sources detected during the first year of the GPSs are listed in Bird et al. (2004). Among the sources, a few tens of them have never been detected prior to {\em INTEGRAL}'s observation. The second year catalog has doubled the number of sources that are either new or without a firm classification (Bird et al. 2006). Most of the new objects from the first catalog share common characteristics such as their location in the Norma region and high intrinsic absorption (Lutovinov et al. 2005). They are believed to be HMXBs (see Kuulkers 2005), and \igr~($l=338.0\degr,b=0.1\degr$) is probably a member of this class. That this system may be a HMXB has also been suggested by Sugizaki et al. (2001) and Combi et al. (2004) due to the huge hydrogen column density towards the source, the hard spectral index (0.7--10 keV band), and its flux variability.

Bodaghee et al. (2006) studied \igr~with {\em INTEGRAL} and {\em XMM-Newton} observations. {\em XMM-Newton EPIC} clearly detected pulsations at 2--10 keV with a 38 $\pm$ 5\% pulse fraction and allowed for a refined position measurement; a potential counterpart, 2MASS J16390535-4642137, is about 2$\arcsec$ away from the position. Using 15--40 keV {\em INTEGRAL ISGRI} data spanning about 54 days, the pulse period was measured to be 912.0 $\pm$ 0.1 s with a pulse fraction of 54 $\pm$ 24\%. By fitting a Comptonized emission model ({\tt comptt} in XSPEC) to the non-simultaneous {\em ISGRI} and {\em EPIC} data, the column density was measured to be $N_{\rm H} = (25 \pm 2) \times 10^{22}$ cm$^{-2}$. The spectrum also required an Fe K$\alpha$ line, and marginally, an Fe K$\beta$ line. The absorbed integrated flux was $4.4 \times 10^{-11}$ erg cm$^{-2}$ s$^{-1}$ in the 2--10 keV band and $5.1 \times 10^{-11}$ erg cm$^{-2}$ s$^{-1}$ in the 20--60 keV band. By creating pulse phase-resolved spectra, Bodaghee et al. (2006) found that the pulsations affect the normalizations but do not modify the spectral shape significantly.

In this work, we use recent and archival {\em RXTE} observations to determine the orbital parameters for \igr~using pulse arrival time analysis. We find three potential solutions to the orbit, although only one is particularly convincing. The observations are described in \S~2. In \S~3 we describe the pulse arrival time measurements and error analysis, and some basic features of the spectrum. The results of the pulse arrival time analysis are presented in \S~4. We discuss the implications of our results in \S~5.

\section{Observations}
Ten {\em RXTE} observations were obtained (Obs. ID 91080) at four day intervals during 2005 October--November (hereafter; epoch 2), with each observation roughly 12 ks in duration (three to four {\em RXTE}~orbits). Four additional 5 ks observations (two {\em RXTE}~orbits) were obtained at 1 and 2 day intervals in 2006 February (epoch 3). In addition, we use an archival observation (Obs. ID 90069) spanning about 2.2 days from 1.2 years earlier (epoch 1). The epoch 1 data was contaminated by the recurring black hole transient 4U 1630-47, which is about 1$\degr$ from \igr~on the sky. The pointing position for epochs 2 and 3 was offset by 0.2$\degr$, excluding 4U 1630-47 from the field-of-view. A summary of these observations is shown in Table 1. Overall, these data extend over 1.56 yr and probe a range of timescales. 

\begin{figure}
\centering
\includegraphics[width=3in]{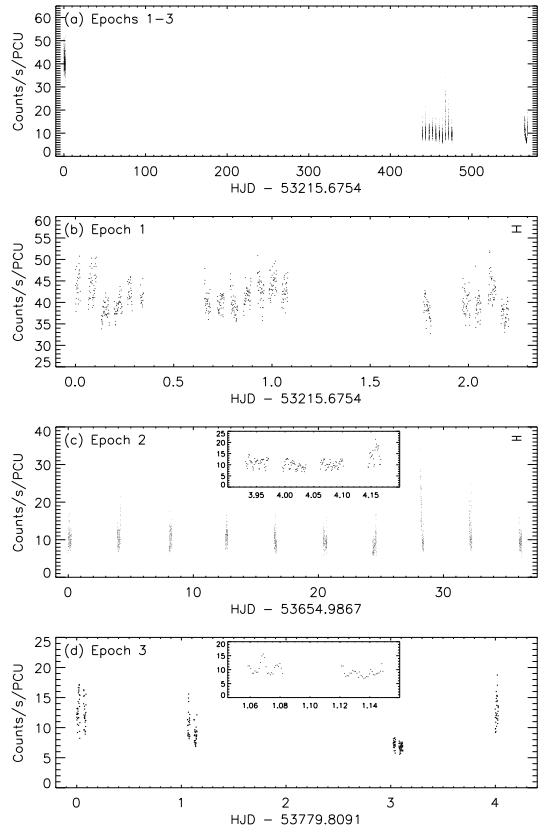}
\caption{Barycenter-corrected and background-subtracted 3--24 keV light curves. From top to bottom the plots are ({\em a}) a merged light curve of all observations, ({\em b}) epoch 1 data, ({\em c}) epoch 2 data, and ({\em d}) epoch 3 data. A characteristic error bar for each epoch is shown in the top-right corner of the panel. The sub-panels for epochs 2 and 3 show variability on shorter time scales. Note the range of time scales present and the higher flux level for the epoch 1 observation due to contamination from 4U 1630-47.\label{alllc}}
\end{figure}

Source and background light curves were created with data from the Proportional Counter Array (PCA; Jahoda et al. 1996) using standard FTOOLS. The PCA instrument consists of five identical multianode proportional counter units (PCUs), operating in the 2--60 keV range, with an effective area of approximately 6500 cm$^{2}$ and a 1$\degr$ field-of-view at FWHM. PCUs 0 and 2 were always operating, but any additional PCUs were also used if their good-time-intervals spanned entire {\em RXTE} orbits. Photon energies were restricted to 3--24 keV, and the light curves were binned by 16 s, which balanced the desires for fine time resolution while maximizing the signal-to-noise. The SkyVLE\footnote{see http.heasarc.nasa.gov/docs/xte/pca\_news.html} background model was used for the epoch 1 data, and the Faint background model was used for epochs 2 and 3. The arrival times of individual events were reduced to the solar system barycenter using the Jet Propulsion Laboratory DE-200 ephemeris (Standish et al. 1992) and {\em faxbary}. The averaged PCU 0 and 2 light curves are shown in Figure \ref{alllc}. Characteristic error bars for each epoch of data are shown in the top-right corner of the corresponding panel. 
\begin{deluxetable*}{cclccc} 
\tablenum{1}
\tabletypesize{\small}
\tablecolumns{5}
\tablewidth{0pt}
\tablecaption{\sc{RXTE Observation Log of IGR J16393-4643}} 
\tablehead{
\colhead{Obs. ID} &
\colhead{Epoch} &
\colhead{Date} &
\colhead{Time Span} &
\colhead{Exposure Time} &
\colhead{Count Rate\tablenotemark{a}} \\
\colhead{} &
\colhead{} & 
\colhead{(U.T.)} &
\colhead{(ks)} &
\colhead{(ks)} &
\colhead{(Counts/s/PCU)} 
}
\startdata
90069-03-02 & 1 & 2004 Jul. 29.67--30.00 & 28.52 & 15.54 & 41.49\tablenotemark{b} \\
90069-03-01 & 1 & 2004 Jul. 30.00--31.99 & 171.94 & 50.22 & 40.77\tablenotemark{b} \\ 
91080-01-01 & 2 & 2005 Oct. 11.98--12.22 & 20.88 & 12.32 & 10.89 \\
91080-01-02 & 2 & 2005 Oct. 15.91--16.17 & 22.08 & 12.50 & 11.50 \\
91080-01-03 & 2 & 2005 Oct. 20.05--20.28 & 20.40 & 11.63 & 11.04 \\
91080-01-04 & 2 & 2005 Oct. 24.56--24.75 & 16.86 & 11.73 & 11.47 \\
91080-01-05 & 2 & 2005 Oct. 28.44--28.66 & 18.96 & 12.19 & 10.48 \\
91080-01-06 & 2 & 2005 Nov. 01.36--01.68 & 27.48 & 12.45 & 10.20 \\
91080-01-07 & 2 & 2005 Nov. 05.30--05.65 & 30.36 & 11.50 & 8.81 \\
91080-01-08 & 2 & 2005 Nov. 09.16--09.41 & 21.30 & 12.38 & 14.70 \\
91080-01-09 & 2 & 2005 Nov. 13.03--13.28 & 21.18 & 12.35 & 12.68 \\
91080-01-10 & 2 & 2005 Nov. 17.03--17.27 & 21.12 & 12.70 & 10.25 \\
91080-01-11 & 3 & 2006 Feb. 13.80--13.90 & 8.48 & 4.69 & 13.60 \\
91080-01-12 & 3 & 2006 Feb. 14.86--14.96 & 8.63 & 4.53 & 10.41 \\
91080-01-13 & 3 & 2006 Feb. 16.83--16.95 & 9.86 & 5.09 & 7.26 \\
91080-01-14 & 3 & 2006 Feb. 17.80--17.91 & 9.26 & 5.28 & 11.68 \\
\enddata
\tablenotetext{a}{For 3--24 keV photons.}
\tablenotetext{b}{Note the contamination during epoch 1 due to 4U 1630-47.}
\end{deluxetable*}

\section{Pulsations}
\begin{figure}
\centering
\includegraphics[width=3.3in]{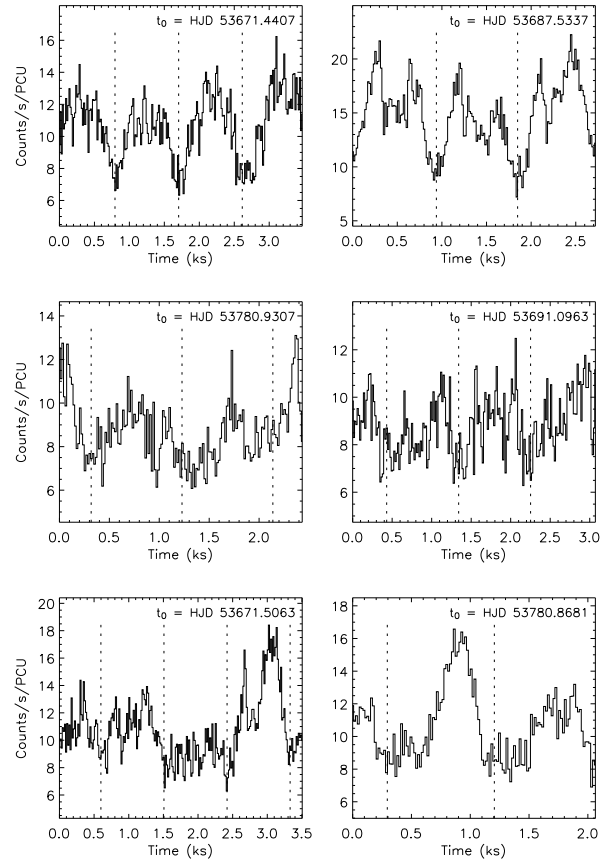}
\caption{Examples of \igr~pulsations using 16 s bins. The top two panels show what are considered to be clean pulsations, the middle two panels show rather noisy or irregular light curve segments, and the lower panels show examples of flares. The pulse arrival times for the six examples are superimposed as vertical dotted lines. Note how the maxima of the flares tend to occur near the maxima of the pulse profile. \label{lcex}}
\end{figure}

 The pulse phase averaged flux varied by $\sim$15\% on hour time scales during epoch 1 (see Fig. \ref{alllc}{\em b}). Due to the contamination from 4U 1630-47, however, the variability cannot unambiguously be attributed to \igr. During epoch 2, the phase averaged flux was constant at the 10\% level for most of the observations, but varied by more than a factor of two over 3 hrs at one point. This outburst was in progress at the beginning of observation 91080-01-08 at HJD 53683.17. The epoch 3 phase averaged flux was also constant within about 10\% for the individual observations (two {\em RXTE}~orbits), but varied by $\sim$80\% over the scale of days. For roughly half of the observations, the pulses are clearly defined, while at other times the light curves are noisy or are highly variable (flares) on $\sim$100 s time scales. Figure \ref{lcex} shows six examples of uninterrupted light curves, each spanning one orbit (hereafter called a ``light segment"). 

\subsection{Arrival Times}
\begin{figure}
\centering
\includegraphics[width=3.3in]{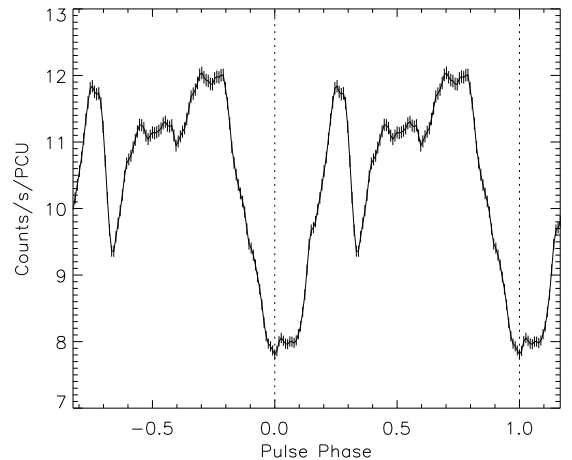}
\caption{Final pulse template (3--24 keV) used to find the pulse arrival times via cross-correlation. The pulse fraction is 21 $\pm$ 1\%. The dotted lines represent the phase corresponding to the pulse arrival times (Table 2). The error bars are 1$\sigma$ statistical uncertainties.\label{pprofile}}
\end{figure}

Pulse arrival times were measured in a multi-step iterative process. We began by removing data within 500 s of a flare, which is defined to be when the flux is more than 75\% above the mean level for the light segment. A preliminary pulse template was then created by manually aligning a selected set of epoch 2 light segments with clearly defined pulses. The alignment was accomplished by fitting a combination of sine functions to the selected light segments, and adjusting the phases accordingly. Manual alignment of the pulses is preferable to folding the entire light curve because it corrects for variations in the arrival times due to the source's position in the orbit. All light segments were folded modulo 912.0 s (Bodaghee et al. 2006), and then the epoch 2 and 3 light segments were cross correlated with the preliminary pulse template. The cross correlation lag times were measured to the nearest second. The resultant pulse arrival times were then used to create a more refined pulse template including all epoch 2 and 3 data (minus the flares). We repeated this process until the pulse arrival times and the pulse template no longer changed. The final pulse template was then used to determine the pulse arrival times for the epoch 1 data separately. In total, sixty pulse arrival times were obtained.\footnote{We do not include two additional pulse arrival times from Bodaghee et al. (2006) due to ambiguity in the location of the pulse minima.} They are presented with their associated errors in Table 2. The final pulse template, with a pulse fraction of 21 $\pm$ 1\%, is shown in Fig. \ref{pprofile}. We note that the pulse template obtained from the PCA is fairly similar to the {\em XMM-Newton} folded light curve (Fig. 7, Bodaghee et al. 2006) with the main and secondary minima occurring at phases 0 and $\sim$0.3.

\subsection{Error Analysis}
Statistical errors on the pulse arrival times were calculated using Monte Carlo simulations. For each light segment, one hundred trial light curves were created. Each point on the light curve was chosen randomly from a Gaussian distribution centered on the measured number of counts, and with the standard deviation determined by Poisson counting statistics from source and background errors added in quadrature. Each trial light curve was cross correlated with the template, yielding slightly different pulse arrival times. The quoted statistical uncertainty is the standard deviation of the difference in pulse arrival time relative to what was measured from each light segment. Statistical errors usually ranged from 3 to 10 s, though the noisier light curves lead to errors of 15 to 30 s, and a few were much larger. 

\begin{figure}
\centering
\includegraphics[width=3in]{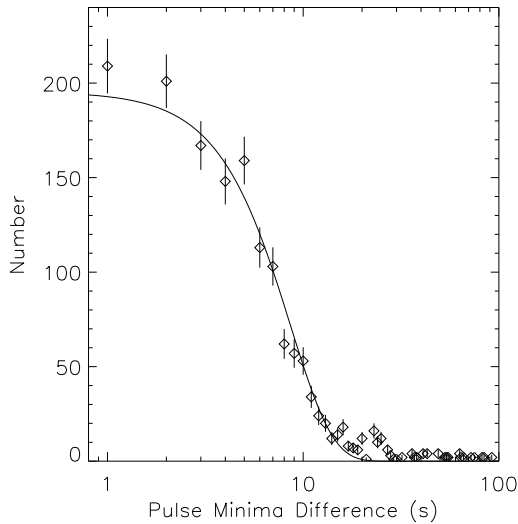}
\caption{Histogram of the difference in pulse minima versus the final template created by using all light segments pulse templates for all other folded light curves. The fitted Gaussian distribution has $\sigma = 5.76$ s. \label{syserrhist}}
\end{figure}

In addition to statistical errors, systematic uncertainties in the pulse arrival times can be caused by distortions in the average pulse profile. These distortions can be caused by varying flux, contamination from nearby sources (e.g., 4U 1630-47), and the energy dependence of the pulse shape can be affected by varying levels of local absorption. To estimate these errors, we aligned each folded light segment with the pulse template using the measured pulse arrival times, and each aligned and folded light segment was then used as a trial pulse template for all of the other folded light segments. The resultant arrival times from each trial generally created a Gaussian distribution centered on the measured arrival time, though occasionally there were outliers. Figure \ref{syserrhist} shows a histogram of the difference in the measured pulse minima versus the final template, created by using all light segments as trial templates for all other light segments, for a total of 1770 total elements. A Gaussian was fitted to the histogram to find a systematic standard deviation of 5.76 s, and this error was added in quadrature to each statistical error. Although the statistical and systematic errors may be correlated, whereby addition in quadrature would overestimate the total errors, the histogram has statistically significant deviations from the normal distribution at a pulse minima difference of 20--30 s, suggesting that the errors associated with some of the pulse arrival times are underestimated in this scenario. Moreover, if the pulse period is different from 912.0 s, folding each light segment on this period will result in slight smearing of the pulsations and will lead to additional systematic errors of order a few seconds.
\subsection{Spectral Analysis}
Motivated by the possibility of orbital modulation of spectral parameters, we extracted one spectrum from each epoch 2 and 3 observation. The epoch 1 spectra were not analyzed due to contamination from 4U 1630-47. The \igr~spectrum consists of emission from the pulsar, plus diffuse emission from the Galactic ridge due to the 1$\degr$ field-of-view of the PCA. The pulsar emission was modeled with an absorbed power-law with a high energy cut-off plus a Gaussian to model iron line emission. The Galactic ridge emission was modeled using the Valinia \& Marshall (1998) parameterization, consisting of an absorbed Raymond-Smith plasma component of temperature $\sim$2--3 keV and a power-law component of photon index $\sim$1.8. The Galactic ridge components were fixed at the Valinia \& Marshall (1998) best-fit values, but we included a variable overall multiplicative constant to adjust the normalization. The fourteen spectra were fitted simultaneously using XSPEC v11.3.2. Unfortunately, we found it difficult to constrain the spectral parameters, and especially the hydrogen column density, due to the Galactic ridge emission. From 3--5 (3--8) keV, the ridge emission accounts for about 75\% (60\%) of the total flux measured by the PCA. This explains the 21 $\pm$ 1\% pulse fraction measured by the PCA, versus the 38 $\pm$ 5\% measured by {\em XMM-Newton} from 2--10 keV, and 54 $\pm$ 24\% by {\em INTEGRAL} from 15--40 keV: The ridge emission effectively washes out the signal. Any uncertainty in the ridge emission normalization leads to even larger uncertainties in the inferred column densities. More importantly, since pulsar spectra are typically complex functions of both phase and energy (White et al. 1983), changes in the \igr~spectrum with pulse phase confuse the analysis.  
\vspace{12.4cm} 
\includegraphics[bb=172 612 432 315,width=3.5in]{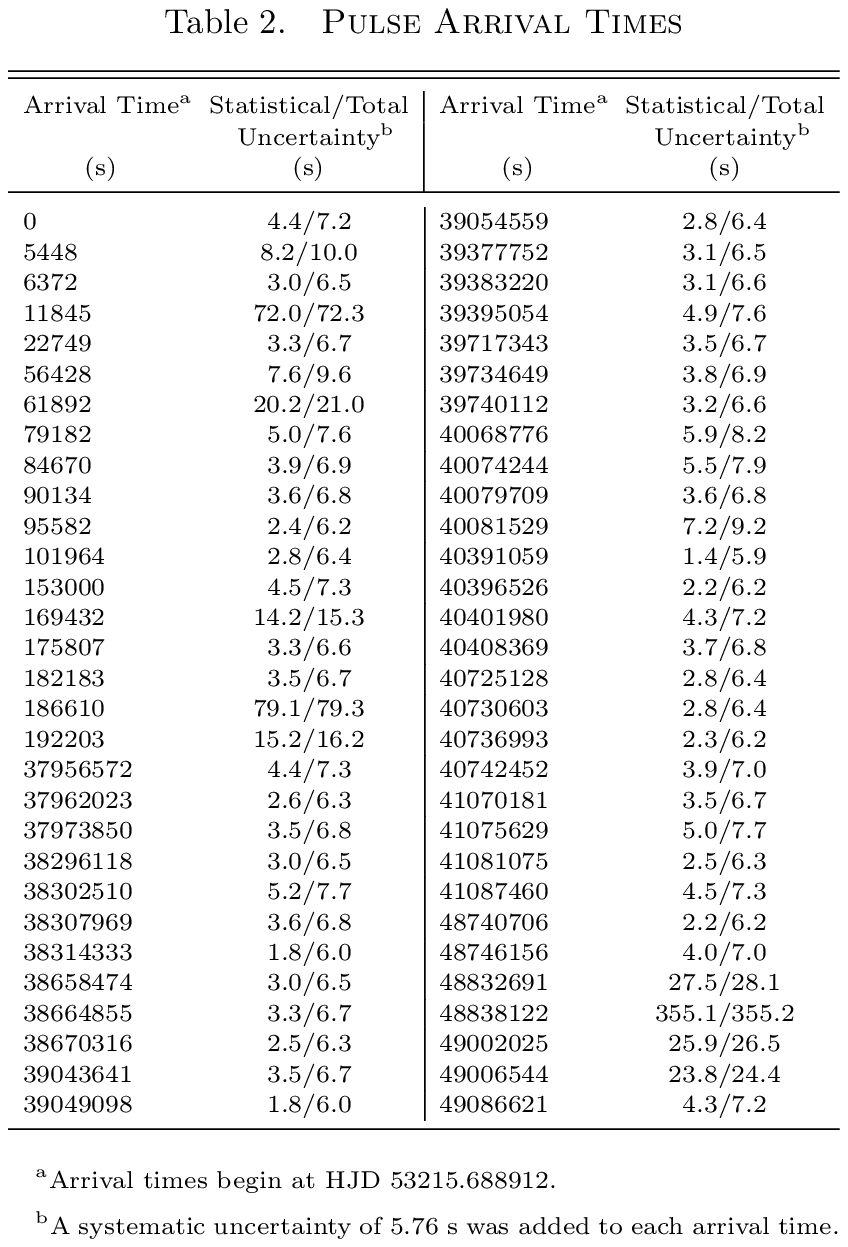}
A final source of confusion in the spectral analysis results from brief flaring episodes, which last for $\sim$100-500 s. Interestingly, the strongest flares usually occurred at the pulse phase corresponding to the maximum of the pulse profile ($\phi \sim 0.7-0.8$, see Fig. \ref{pprofile}). To illustrate spectral distortion during the flares, we created a spectrum of the central 150 s of the flare seen in the lower-right panel of Fig. \ref{lcex} (Obs. ID 91080-01-12-00), and compared it to the spectrum from the remaining portions of that {\em RXTE} orbit. The spectra were fitted simultaneously, with the hydrogen column density and Galactic ridge emission normalization tied to common values. For this particular observation, a high energy cut-off was unnecessary, and $N_{\rm H}=15^{+2}_{-3} \times 10^{22}$ cm$^{-2}$ (90\% confidence interval). The spectrum during the flare was best fit with a photon index $\Gamma = 0.86 \pm 0.14$, while $\Gamma = 1.46^{+0.16}_{-0.12}$ for the rest of the observation. This type of spectral hardening is not unexpected, yet it precludes constraining the hydrogen column density and power-law components simultaneously due to correlations between these parameters. We leave a full phase resolved spectral analysis as future work.

\section{Results}
\begin{figure}
\centering
\includegraphics[width=3in]{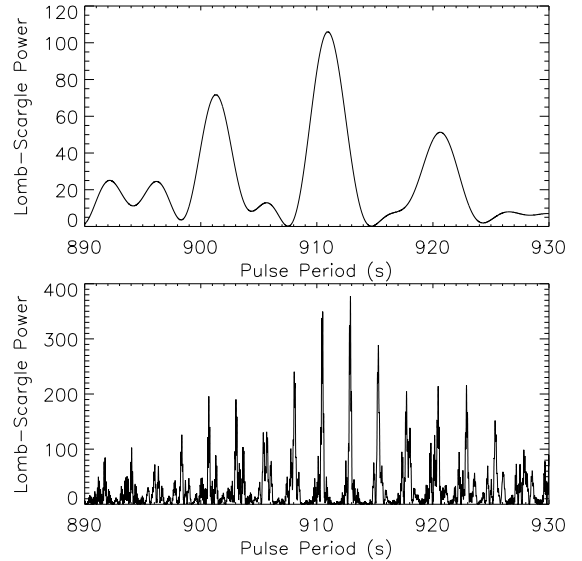}
\caption{Lomb-Scargle periodograms for the epoch 1 data ({\em top}) and for the data from epochs 2 and 3 ({\em bottom}). Note the aliasing of the peaks in the lower panel due to the $\sim$4 day separation of the epoch 2 observations.\label{ls}}
\end{figure}

\begin{figure}
\centering
\includegraphics[width=3in]{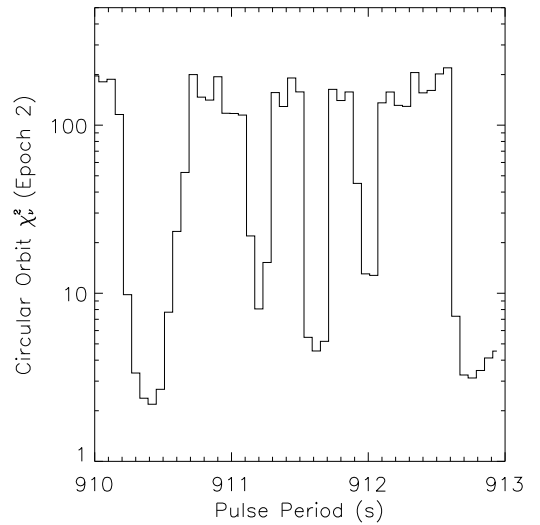}
\caption{Lowest $\chi^{2}_{\nu}$ for all circular orbits to the epoch 2 data as a function of the pulse period (0.06 s bins). The pulse period was allowed to vary by 0.03 s in either direction, and its derivative was allowed to span the range $\mbox{\.{\em P}} = [-3,3] \times 10^{-8}$ s s$^{-1}$.\label{chifits}}
\end{figure}

\begin{figure}
\centering
\includegraphics[width=3.65in]{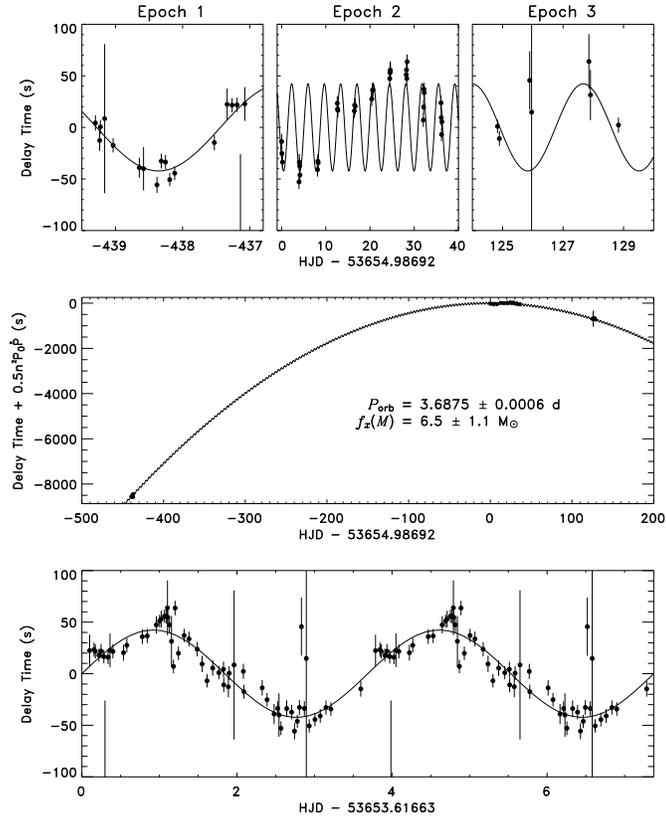}
\caption{Orbital model fit to solution 1. {\em Top panels:} Zoomed in plots of the fits to each epoch. {\em Middle panel:} Fit to all epochs, explicitly showing the change in the pulse period over $\sim$1.5 yr. {\em Bottom panel:} All epochs plotted modulo the orbital period for two cycles.\label{fit9104s}}
\end{figure}

\begin{figure}
\centering
\includegraphics[width=3.65in]{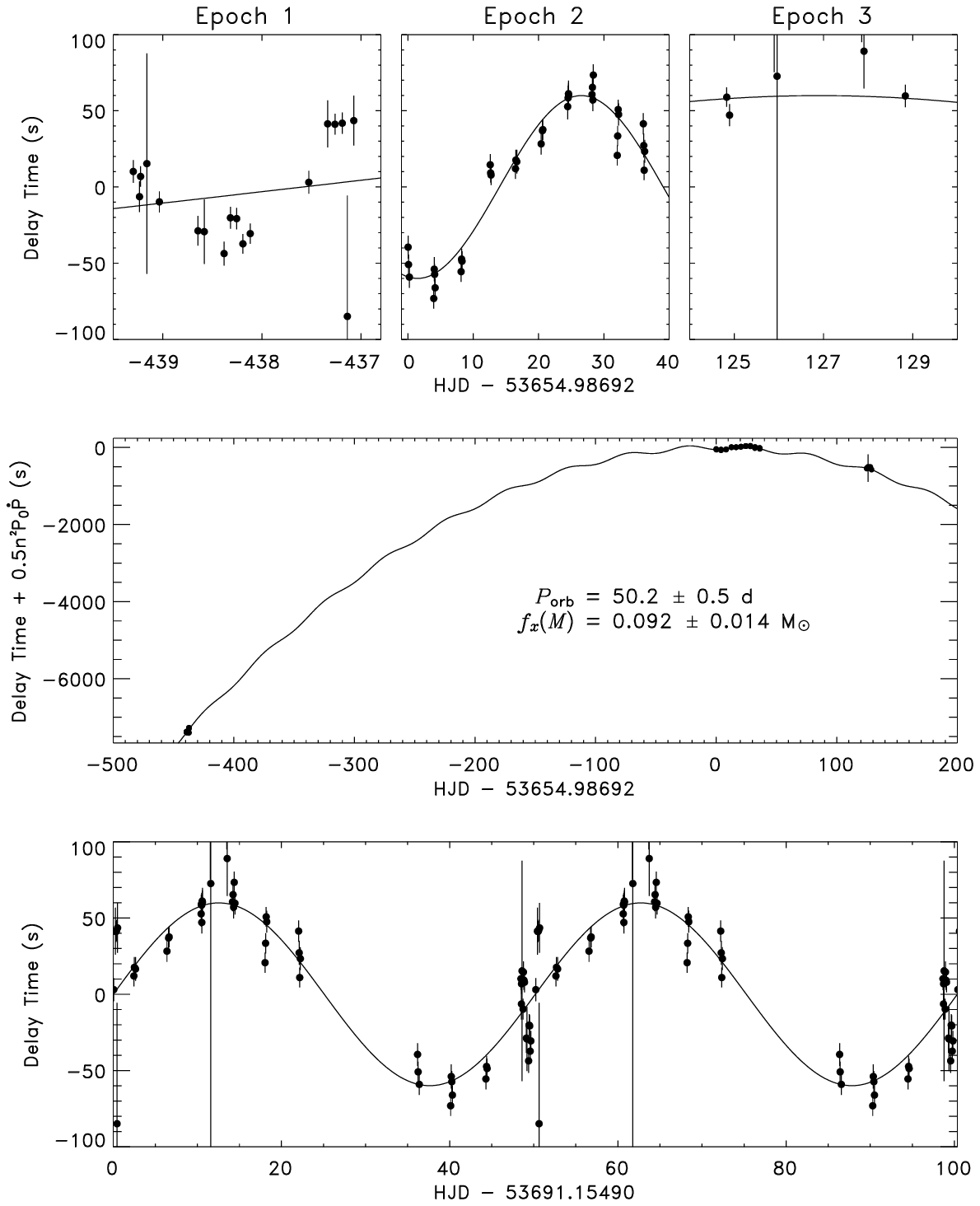}
\caption{Same as Fig. \ref{fit9104s}, but for orbital solution 2.\label{fit9104l}}
\end{figure}

\begin{figure}
\centering
\includegraphics[width=3.65in]{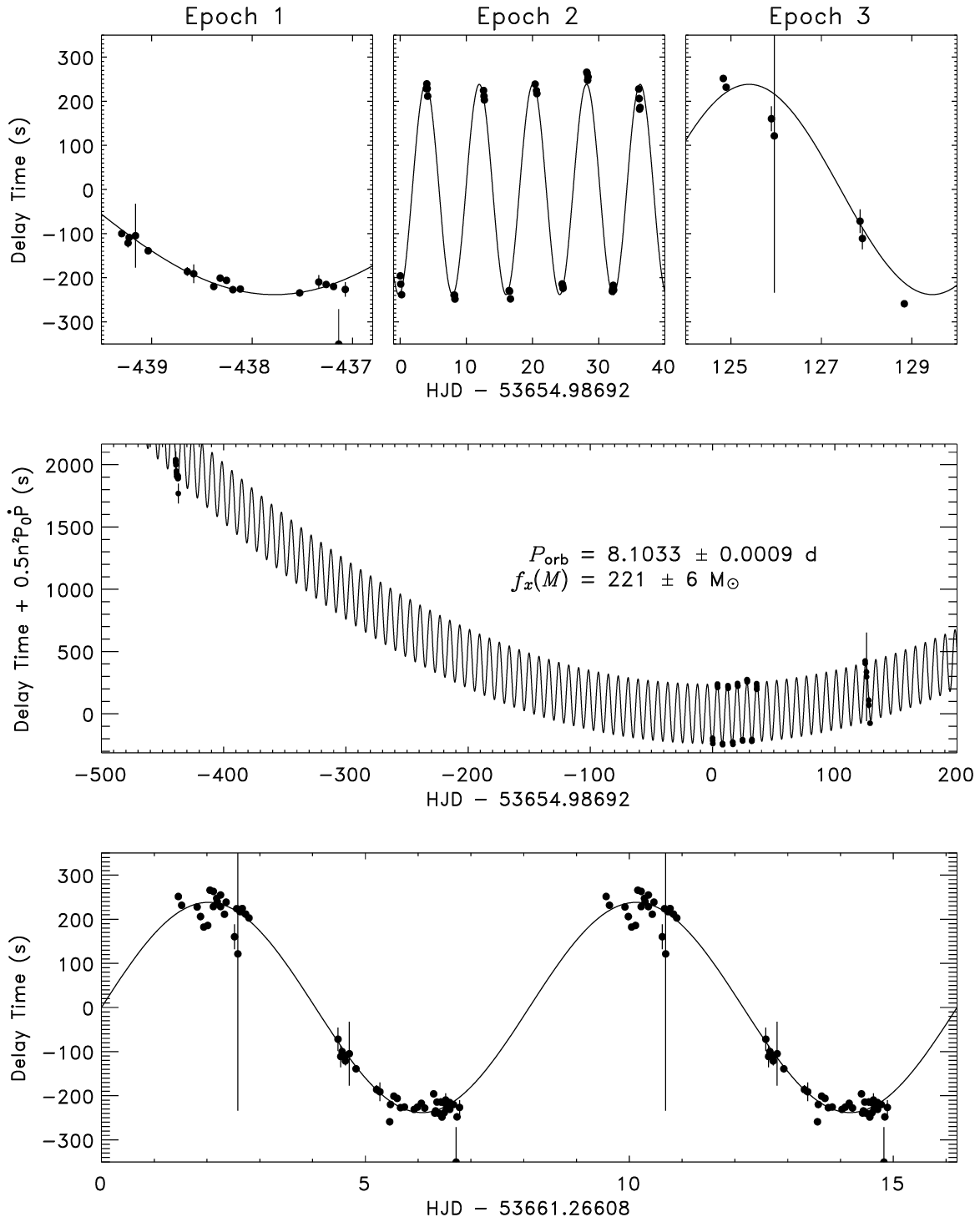}
\caption{Same as Fig. \ref{fit9104s}, but for orbital solution 3.\label{fit9116}}
\end{figure}

The pulse arrival times were fitted with a seven parameter model of the orbit and pulse period evolution. The arrival time of the $n$th pulse is given by
\begin{equation} 
t_{n}=t_{0}+nP_{0}+\frac{n}{2}P_{0}  \mbox{\.{\em P}} + a_{x}\sin{i}\sin{\left[\frac{2\pi (t_{n}-t_{0})}{P_{\rm orb}}+\phi_{0}\right]},
\end{equation}
where $P_{0}$ is the pulse period at time $t_{0}$, $\mbox{\.{\em P}}$ is the pulse period derivative and is assumed to be constant, $a_{x}\sin{i}$ is the projected semi-major axis of the orbit, $P_{\rm orb}$ is the orbital period of the system, and $\phi_{0}$ is the phase of the orbit at time $t_{0}$. The pulse number $n$ is given by the nearest integer to
\begin{equation}
n=\frac{t_{n}-t_{0}}{\langle P \rangle}=\frac{t_{n}-t_{0}}{P_{0}+0.5 \mbox{\.{\em P}}(t_{n}-t_{0})}.
\end{equation}
We made sure the pulse period had not substantially changed from 912.0 s by creating Lomb-Scargle periodograms for the epoch 1 and epoch 2/3 data separately, which are presented in Figure \ref{ls}. Determination of the orbital elements is complicated by the need to decouple the effects of the orbital Doppler delays and the intrinsic changes in the neutron star rotation rate. To minimize these concerns, we first searched for circular orbital solutions using the epoch 2 data only. The pulse arrival times were fitted to equation (1), using a fine grid of initial values for the pulse period, projected semi-major axis, orbital period, and phase. An extensive search led to five potential fits that can be seen as the minima in Fig. \ref{chifits}, which shows the lowest $\chi^{2}_{\nu}$ for circular orbits fits to the epoch 2 data as a function of pulse period. Given these fits, we then utilized all data by phase connecting the epoch 2 data to the data from epochs 1 and 3. To connect the three epochs of data, we used the best-fit circular orbit parameters for the epoch 2 data as initial values and cycled through a grid of pulse period derivatives $[-3,3] \times 10^{-8}$ s s$^{-1}$ with a step-size of $5 \times 10^{-10}$ s s$^{-1}$. One of the initial fits was rejected due to an inability to connect the data sets, and one was rejected due to a substantially larger value of $\chi^{2}_{\nu}$, leaving three potential solutions. The results of the fits are presented in Table \ref{fits}, and the orbital pulse arrival delays are plotted in Figures \ref{fit9104s}--\ref{fit9116}. The error bars in the figures represent the 68\% confidence level, and the secular change in the pulse periods has been removed from the top and bottom panels so that the modulation is purely due to orbital motion.
\begin{deluxetable*}{lcccc} 
\tablenum{3}
\tabletypesize{\small}
\tablecolumns{5}
\tablewidth{0pt}
\tablecaption{\sc{Circular Orbital Solutions to All Epochs}\label{fits}} 
\tablehead{
\colhead{Solution:} & 
\colhead{} &
\colhead{1} & 
\colhead{2} & 
\colhead{3}  
}
\startdata
$P_{\rm pulse}$\tablenotemark{a} & (s) & 910.4178 $\pm$ 0.0006 & 910.4028 $\pm$ 0.0007 & 911.6167 $\pm$ 0.0005  \\
$\mbox{\.{\em P}}$ & ($\times 10^{-9}$ s s$^{-1}$) & -10.81 $\pm$ 0.03 & -9.34 $\pm$ 0.04 & 2.7 $\pm$ 0.3 \\
$a_{x}\sin{i}$ & (lt-s) & 43 $\pm$ 2 & 60 $\pm$ 3 & 238 $\pm$ 2 \\
$P_{\rm orb}$ & (d) & 3.6875 $\pm$ 0.0006 & 50.2 $\pm$ 0.5 & 8.1033 $\pm$ 0.0009 \\
$f_{x}(M)$ & $({\rm M}_{\sun})$ & 6.5 $\pm$ 1.1 & 0.092 $\pm$ 0.014 & 221 $\pm$ 6  \\
$\chi^2_{\nu}$ & & 2.55 & 5.29 & 7.72 \\
$\chi^2_{\nu},{\rm rnm}$\tablenotemark{b} & & 1.00 & 2.07 & 3.02 \\
\enddata
\tablecomments{All errors are quoted at the 90\% confidence level for a single parameter. The fits to eq. (1) each have 54 degrees of freedom.}
\tablenotetext{a}{Pulse periods at $t_{0} = {\rm HJD}~53654.98692$.}
\tablenotetext{b}{The renormalized $\chi^2_{\nu}$ was obtained by multiplying each total error by 1.6.} 
\end{deluxetable*}
Of the three potential orbital solutions presented in Table \ref{fits}, only orbital solution 1 appears convincing. This solution has a projected semi-major axis of 43 $\pm$ 2 lt-s and an orbital period of 3.6875 $\pm$ 0.0006 days. The X-ray mass function is 
\begin{equation}
f_{x}(M) = \frac{(M_{c} \sin{i})^{3}}{(M_{x}+M_{c})^{2}} = \frac{4 \pi^2 (a_{x}\sin{i})^{3}}{GP_{\rm orb}^{2}} = 6.5 \pm 1.1~{\rm M}_{\sun},
\end{equation}
which is consistent with the HMXB interpretation proposed by Sugizaki et al. (2001), Combi et al. (2004), and Bodaghee et al. (2006). Finally, a pulse period of 910.4 s with a derivative of $-1.08 \times 10^{-8}$ s s$^{-1}$ implies a pulse period of about 911.3 $\pm$ 0.1 s at the time of the 912.0 $\pm$ 0.1 s pulse period determination of Bodaghee et al. (2006). The 5$\sigma$ difference potentially indicates that the pulse period derivative was not constant. This most likely was due to the average accretion rate increasing between 2003 February--March and 2004 July.  

Orbital solution 2 is statistically acceptable, but it is not convincing for four reasons: (1) The value of $\chi^{2}_{\nu}$ (see Table \ref{fits}) is twice as large as that of orbital solution 1. (2) The residuals to the phase connection to the epoch 1 data systematically deviate from the orbital model. (3) There is a clear spread in the three to four pulse minima from each day of epoch 2 data, implying the pulse period and orbital period are not compatible. (4) Finally, with a mass function of 0.09 M$_{\sun}$, the system would have to be viewed nearly face-on ($i < 13\degr$) for the companion to be a high mass star (assuming a 1.4 M$_{\sun}$ pulsar and a companion mass of 10 M$_{\sun}$). We also note that about 30\% of the orbital phase is not sampled with these data. 

Orbital solution 3 is immediately suspect because its mass function is greater than 200 M$_{\sun}$, and a star of this mass is hard (if not impossible) to rectify with theoretical predictions for the largest possible stellar mass (e.g., Krumholz 2005). If confirmed, solution 3 would represent an extremely important discovery because the orbital companion would more likely be an intermediate mass black hole (IMBH) rather than a normal star. Since the pulsar is observable in X-rays there is obviously a mass-donating companion, and therefore the system would necessarily be a triple system, which begs the question of how such a system could be dynamically formed. Moreover, there may be problems fitting the orbit of the pulsar and mass-donating companion outside of the Roche lobe of the IMBH. Unusually large companion mass aside, orbital solution 3 can be phase-connected over all three epochs, and a second orbit added to eq. (1) could possibly bring $\chi^{2}_{\nu}$ to a level comparable to the other solutions. On the other hand, the secular pulse period derivative is positive, which is clearly inconsistent with the 912.0 $\pm$ 0.1 s pulse period measurement. Similar to orbital solution 2, nearly half of the orbital phase is not sampled with these data. 

Orbital solution 1 stands out as the most likely one, yet with $\chi^2_{\nu}=2.55$, the errors associated with the pulse arrival times would necessarily be underestimated by an average of about 60\% to yield $\chi^2_{\nu}=1.00$. This is not unreasonable considering the quality of the data (see Fig. \ref{lcex}), and the fact that there are statistically significant deviations from the Gaussian fit to the histogram used to estimate the systematic errors. As a final attempt to improve the fit, we added eccentricity to the orbital model by including the term 
\begin{equation}
-\frac{e}{2}a_{x}\sin{i}\sin{\left[\frac{4\pi (t_{n}-t_{0})}{P_{\rm orb}}+\phi_{0}-\omega_{\rm p}\right]} 
\end{equation}
in eq. (1), where $e$ is the eccentricity and $\omega_{\rm p}$ is the longitude of periastron. The additional term represents the first-order term in a Taylor series expansion in the eccentricity and is reasonably accurate for a mildly eccentric orbit (Levine et al. 2004). An F-test showed that the addition of eccentricity to orbital models 2 and 3 was not significant, with chance probabilities of 0.16 and 0.09. Orbital solution 1, on the other hand, improved with a 3.4\% probability of the improvement occuring by chance ($\chi^{2}/\nu\mid_{{\rm circ} \rightarrow {\rm ecc}}=  137.7/54 \rightarrow 120.9/52$). With the improved fit to orbital solution 1, the best-fit parameters remained approximately constant with the exception of the projected semi-major axis which increased from $43 \pm 2$ to $55 \pm 2$ lt-s, yielding a new mass function of $13.1 \pm 1.0$ M$_{\sun}$. The best-fit eccentricity was $0.15 \pm 0.05$. 

\section{Discussion}
By all indications, \igr~is a HMXB system. It is heavily absorbed, with roughly an order of magnitude more absorption than can be attributed to the Galactic interstellar medium (Dickey \& Lockman 1990), indicating intrinsic absorption within the binary system. Such strong absorption is probably due to the stellar wind of a young companion star. Low-mass X-ray binaries, on the other hand, are generally not heavily extincted like \igr. In addition, sources like \igr~that undergo occasional flaring are typically found in wind-fed systems. Still, the nature of the companion star for this source is not certain. The most likely orbital solution has an X-ray mass function of about $6.5~{\rm M}_{\sun}$, although the reader should bear in mind that including mild eccentricity to the orbit increases the mass function to $13.1 \pm 1.0$ M$_{\sun}$. Nevertheless, it's position on the plot of pulse period versus orbital period---the so-called Corbet diagram (Corbet 1984)---implies that the companion is a supergiant with an underfilled Roche lobe rather than a Be star. The absence of transient behavior and the probable circular orbit further suggest that this is a supergiant system and not a Be X-ray binary. Until recently, only 10 supergiant X-ray binaries were known to exist in the Galaxy (e.g., Liu et al. 2000), although many new members of this class are being uncovered by the {\em INTEGRAL} GPSs (Walter et al. 2006). Eclipses are often observed in pulsars with supergiant binary companions if the inclination of the system is high enough. By folding the PCA light curves modulo the orbital period of the three solutions, we do not find any evidence for eclipses, which allows the constraint $R_{c} \tan{i} \la 18.5~{\rm R_{\sun}}$ if we assume the companion mass is much larger than that of the pulsar. Assuming a companion radius of 10 R$_{\sun}$, the inclination must be less than about 60$\degr$. Nevertheless, further observations are required to obtain an unambiguous orbital solution for \igr. When combined with an optical mass function and a constraint on the inclination, the X-ray mass function would determine the masses of the two stars, which is important for constraining neutron star equations of state, and for understanding the evolution of binary systems.

\acknowledgements
We would like to thank R\"{u}diger Staubert for providing insight into the nature of the problem at the early stages of this work, and Jean Swank and Evan Smith for help in planning the {\em RXTE} observations. This research has made use of data obtained through the High Energy Astrophysics Science Archive Research Center Online Service, provided by the NASA/Goddard Space Flight Center.


\begin{thebibliography}{}
\bibitem[]{}Bird, A. J., Barlow, E. J., Bassani, L., et al. 2004, \apj, 607, L33
\bibitem[]{}Bird, A. J., Barlow, E. J., Bassani, L., et al. 2006, \apj, 636, 765
\bibitem[]{}Bodaghee, A., et al. 2006, \aap, 447, 1027 
\bibitem[]{}Combi, J. A., Ribo, M., Mirabel, I. F., \& Sugizaki, M. 2004, \aap, 422, 1031
\bibitem[]{}Corbet, R. H. D. 1984, \aap, 141, 91
\bibitem[]{}Dickey, J. M., \& Lockman, F. J. 1990, \araa, 28, 215
\bibitem[]{}Jahoda, K., Swank, J., Giles, A. B., Stark, M. J., Strohmayer, T., \& Zhang, W. 1996, Proc. SPIE, 2808, 59
\bibitem[]{}Krumholz, M. R. 2005, in ASP Conf. Ser., Frank N. Bash Symposium 2005: New Horizons in Astronomy, eds. Kannappan, S., Redfield, S., Drory, N., Kessler-Silacci, J., \& Landriau, M. (San Francisco: ASP), in press
\bibitem[]{}Kuulkers, E. 2005, in Interacting Binaries: Accretion, Evolution and Outcomes, eds. L.A. Antonelli, et al. (AIP: New York), 402
\bibitem[]{}Levine, A. M., Rappaport, S., Remillard, R., \& Savcheva, A. 2004, \apj, 617, 1284
\bibitem[]{}Liu, Q. Z., van Paradijs, J., \& van den Heuvel, E. P. J. 2000, A\&AS, 147, 25
\bibitem[]{}Lutovinov, A., Revnivtsec, M., Gilfanov, M., Shykovskiym, P., Molkov, S., \& Sunyaev, R. 2005, \aap, 444, 821 
\bibitem[]{}Standish, E. M., Newhall, X. X., Williams, J. G., \& Yeomans, D. K. 1992, in Explanatory Supplement to the Astronomical Almanac, ed. P. K. Seidelmann (Mill Valley: University Science), 279
\bibitem[]{}Sugizaki, M., Mitsuda, K., \& Kaneda H., et al. 2001, ApJS, 134, 77
\bibitem[]{}Valinia, A., \& Marshall, F. E. 1998, \apj, 505, 134
\bibitem[]{}Walter, R., et al. 2006, \aap, in press
\bibitem[]{}White, N. E., Swank, J. H., \& Holt, S. S. 1983, \apj, 270, 711
\end{thebibliography}
\end{document}